\documentstyle[aps,amssymb,preprint,psfig]{revtex}
\tightenlines
\begin{document}
\title{Entanglement and visibility at the output of a Mach-Zehnder
interferometer}
\author{Matteo G. A. Paris}
\address{{\sc Theoretical Quantum Optics Group} \\ 
Dipartimento di Fisica 'Alessandro Volta' dell'Universit\'a di Pavia \\
Istituto Nazionale di Fisica della Materia -- Unit\'a di Pavia\\ 
via Bassi 6, I-27100 Pavia, ITALY}
\date{Accepted for publication on PRA: 04Nov98}
\maketitle
\begin{abstract}
We study the entanglement between the two beams exiting a Mach-Zehnder
interferometer fed by a couple of squeezed-coherent states with arbitrary 
squeezing parameter. The quantum correlations at the output are function 
of the internal phase-shift of the interferometer, with the output state 
ranging from a totally disentangled state to a state whose degree of 
entanglement is an increasing function of the input squeezing parameter. 
A couple of squeezed vacuum at the input leads to maximum entangled state 
at the output. The fringes visibilities resulting from measuring the 
coincidence counting rate or the squared difference photocurrent are evaluated 
and compared each other. Homodyne-like detection turns out to be preferable 
in almost all situations, with the exception of the very low signals regime. 
\end{abstract}
\section{Introduction}\label{s:intro}
The notion of entanglement is an essential feature of quantum mechanics, 
and is strictly connected with the nonlocal character of the theory. 
A two-part physical system prepared in an entangled state is described 
by a non factorizable density matrix. This gives raise to partial or total 
correlation between the outcomes of measurements performed on the two
parts, even though the parts may be so far apart that no effects 
resulting from one measurement can reach the other part within the 
light cone.  
\par
Sources of entangled states are required for fundamentals tests of quantum 
mechanics, as well as for applications such as quantum computation 
and communication \cite{be95}, and teleportation \cite{tdm,tbr}. 
In recent years, entangled photon pairs had been used to test non-locality 
of quantum mechanics \cite{bl1,bl2,bl3,bl4} by Bell inequality \cite{bell}. In 
practice, all the available sources of two-mode entangled states are based 
on the process of spontaneous down-conversion, taking place in $\chi^{(2)}$ 
nonlinear crystals \cite{sdc1}. Recently, it has been 
demonstrated that a beam splitter can split an incident photon into two 
correlated secondary photons \cite{fra1,fra2}. However, such process occurs 
at very low rate, and thus it is of no interest in practical applications. 
\par
In order to study quantum correlations between two radiation modes, and 
to compare different sources of correlated states, one needs to quantify 
the degree of entanglement \cite{bp91}. A good theoretical measure of 
correlations has been introduced by means of Von-Neumann entropy. The 
entropy of a 
two-mode state $\hat\varrho$ is defined as 
\begin{eqnarray}
S[\hat\varrho] = - \hbox{Tr}\left\{\hat\varrho \log\hat\varrho\right\}
\label{totS}\;,
\end{eqnarray}
whereas the entropies of the two modes $a$ and $b$ are given by
\begin{eqnarray}
S[\hat\varrho_a] = -\hbox{Tr}_a\left\{\hat\varrho_a \log\hat\varrho_a\right\} 
\qquad S[\hat\varrho_b] = - \hbox{Tr}_b\left\{\hat\varrho_b \log\hat
\varrho_b\right\}
\label{abS}\;.
\end{eqnarray}
In Eq. (\ref{abS}) $\hat\varrho_a = \hbox{Tr}_b\left\{\hat\varrho\right\}$
and $\hat\varrho_b = \hbox{Tr}_a\left\{\hat\varrho\right\}$ denote the
state of $a$ and $b$ respectively, as obtained by tracing out the other mode
from the total density matrix. Following Refs. \cite{bp91,lind,bp89,ve97} 
we define the degree of entanglement of the state $\hat\varrho$ as the 
normalized excess entropy \cite{note}
\begin{eqnarray}
\epsilon = \frac{1}{S[\hat\varrho_{a{\sc th}}]+ S[\hat\varrho_{b {\sc th}}]} 
\Bigg\{ S[\hat\varrho_a] + S[\hat\varrho_b] - S[\hat\varrho] \Bigg\}
\label{entdef}\;,
\end{eqnarray}
where $S[\hat\varrho_{\sc th}]=\log (1+N) + N \log (1+N^{-1})$, with $N=
\langle a^\dag a\rangle$, denotes the entropy of a thermal state, namely
the maximally disordered state at fixed intensity.  The use of
$\epsilon$ formalizes the idea that the stronger are the correlations
in the two-mode state, the more disordered should be the two modes taken
separately. If $\hat\varrho$ is a pure state, we have that $S=0$ and
$S_a=S_b$ \cite{arak}, so that $\epsilon=S[\hat\varrho_a]/ 
S[\hat\varrho_{a \sc th}]$ ranges from zero to unit. Notice that for pure 
state $\epsilon$ represents the unique measure of entanglement \cite{po97}.
\par
From the experimental point of view, the entanglement can be detected by
non-classical interference effects occurring in intensity-dependent
measurements. In experiments involving photon pairs from parametric
down-conversion these effects occur when coincident photons are mixed
at a beam splitter \cite{ho87,ma91,ra92}.  
The probability amplitudes for pairs from the two
arms show destructive interference, leading to a suppression of the
coincidence counting rate between detectors surveying the two arms 
\cite{fe87,fe89}. Recently, the spatial effects in two-beam 
interference have been also studied for partially entangled photon pairs 
\cite{sa98}.  
In the case of more excited states, many photons are present  
and the connection between entanglement and coincidence rate is less
transparent. The issue has been received attention \cite{bern,di97}, 
though a general theory has not been developed yet. 
\par
In this paper we study the generation and the detection of entangled
states at the output of a Mach-Zehnder interferometer fed by a couple of
uncorrelated squeezed-coherent states. The scheme may be of interest as
the output state can be arbitrarily high excited, instead of having two
photons only. In addition, the degree of entanglement can be tuned by
varying the degree of squeezing of the input beams, or the internal
phase shift of the interferometer.
As regards the detection scheme, we show that the coincidence counting
rate between the two output arms corresponds to low fringes visibility.
Therefore, we consider instead another intensity dependent quantity, namely 
the squared difference photocurrent, which shows high visibility of fringes
for the whole range of input squeezing parameter. 
\par
In Section \ref{s:ent} we study the dynamics of the interferometer, and
evaluate analytically the degree of entanglement at the output as a 
function of the squeezing fraction of the input beams and the internal 
phase-shift of the interferometer. In Section \ref{s:vis} we analyze the 
interference effects occurring in the measurement of the photon coincidence 
rate and of the squared difference photocurrent. The evaluation of the 
fringes visibility for both measurements shows that homodyne-like detection 
is preferable in almost all situations, with the exception of the very low
signals regime. Section \ref{s:outro} closes the paper with some
concluding remarks.
\section{Entanglement at the output of a Mach-Zehnder interferometer}
\label{s:ent}
\begin{figure}
\psfig{file=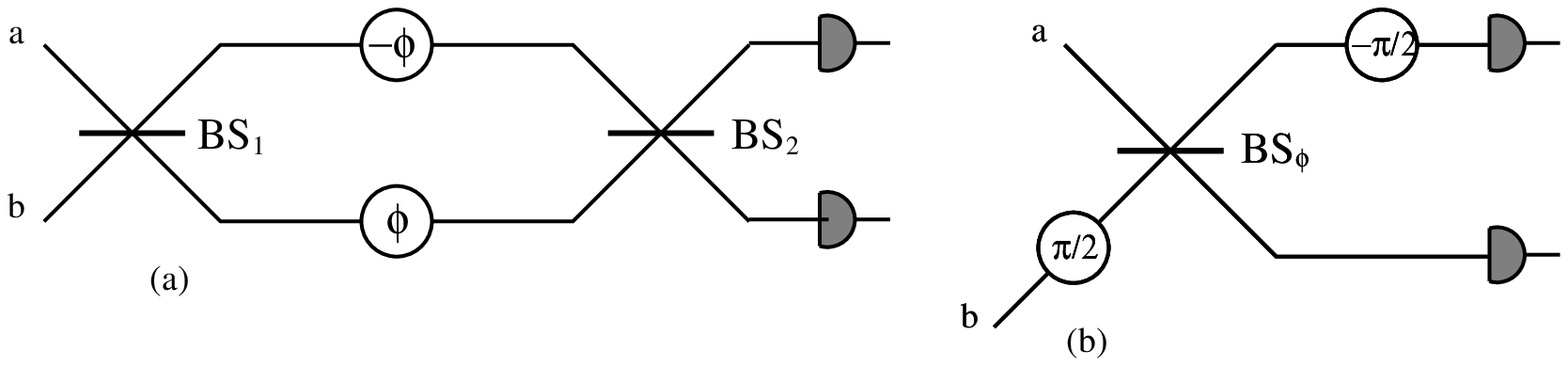,width=14cm} 
\caption{In (a): schematic diagram of a Mach-Zehnder interferometer.
$BS_1$ and $BS_2$ are symmetric beam splitters, whereas $a$ and $b$
denote the input signal modes. Equal and opposite phase-shifts are
imposed in each arm. In (b): an equivalent scheme for the Mach-Zehnder
interferometer depicted in (a): a single beam splitter $BS_\phi$ of
transmissivity $\tau=\cos^2 \frac{\phi}{2}$ preceded and followed by
rotations of $\frac{\pi}{2}$ performed on one of the two modes (here
$b$). \label{f:sch}}
\end{figure}

The Mach-Zehnder interferometer we are dealing with is depicted in Fig.
\ref{f:sch}a. The input signal modes are denoted by $a$ and $b$, whereas
$BS_1$ and $BS_2$ are symmetric beam splitters. We also assume that
equal and opposite phase-shifts $\phi$ are imposed in each arm of the
interferometer.  The evolution operator of the whole setup can be
written as 
\begin{eqnarray}
\hat V_{\small MZ}(\phi)= \hat Ue^{i\phi (a^\dag a-b^\dag b)}\hat U^\dag 
\label{mzevol1}\;,
\end{eqnarray}
where 
\begin{eqnarray}
\hat U =\exp \left\{ i \frac{\pi}{4} (a^\dag b + b^\dag a)\right\} 
\label{50bs}\;,
\end{eqnarray}
denotes the evolution operator of a symmetric beam splitter.
After straightforward algebra one rewrites Eq. (\ref{mzevol1}) as
\begin{eqnarray}
\hat V_{\small MZ}(\phi)=\exp\left\{ i \frac{\pi}{2} b^\dag b\right\}
\exp\left\{-i\frac{\phi}{2}(a^\dag b+b^\dag a)\right\}
\exp\left\{-i\frac{\pi}{2}b^\dag b\right\}
\label{mzevol2}\;,
\end{eqnarray}
which shows that a Mach-Zehnder interferometer is equivalent to a single 
beam splitter $BS_\phi$ of transmissivity $\tau=\cos^2 \frac{\phi}{2}$, 
preceded and followed by rotations of $\pi/2$ performed on one of the 
two modes (see Fig. \ref{f:sch}b). 
\par\noindent
We consider the Mach-Zehnder interferometer fed by a couple of 
squeezed-coherent states
\begin{eqnarray}
|\psi_{\small \sc in}\rangle\hat D_a (\alpha) \hat D_b (\alpha) 
\hat S_a (\zeta ) \hat S_b (\zeta )| {\bf 0} \rangle =
\label{psin}\;.
\end{eqnarray}
In Eq. (\ref{psin}) $\hat D_a (\alpha)=\exp (\alpha a^\dag - 
\bar{\alpha} a)$ is the displacement operator and $\hat S (\zeta) =
\exp[1/2(\zeta^2 a^{\dag 2} - \bar{\zeta}^2 a^2)]$ is the squeezing 
operator, $| {\bf 0} \rangle$ denotes the electromagnetic vacuum.  
There is no need to consider a phase-shift between the input modes, as 
it can be reabsorbed into the internal phase shift $\phi$.
Without loss of generality, in the following we will consider a complex 
field amplitude $\alpha \in {\Bbb C}$ and a real squeezing parameter
$\zeta\equiv r \in {\Bbb R}$. 
\par\noindent
The state exiting the interferometer is given by
\begin{eqnarray}
|\psi_{\small \sc out}\rangle= \hat V_{\small MZ} (\phi)\:|
\psi_{\small \sc in}\rangle 
\label{outmz0}\;.
\end{eqnarray}
By exploiting the vacuum invariance $\hat V_{\small 
MZ} (\phi) |{\bf 0}\rangle = 
|{\bf 0}\rangle$ and using the relation
\begin{eqnarray}
\exp\left\{-i\frac{\pi}{2} a^\dag a\right\} 
\hat D (\alpha) \hat S (r)
\exp\left\{i\frac{\pi}{2} a^\dag a\right\} 
=\hat D (-i\alpha) \hat S (-r) 
\label{form}\;,
\end{eqnarray}
we can write $|\psi_{\small \sc out}\rangle$ as
\begin{eqnarray}
|\psi_{\small \sc out}\rangle =\exp\left\{i \frac{\pi}{2} b^\dag b\right\} \:\:
\hat U_\phi \hat D_a (\alpha) \hat D_b (-i\alpha) \hat U_\phi^\dag \:\:
\hat U_\phi \hat S_a (r) \hat S_b (-r)\hat U_\phi^\dag \:\:
|{\bf 0}\rangle
\label{outmz1}\;,
\end{eqnarray}
where $\hat U_\phi$ denotes the evolution operator of the equivalent beam
splitter $BS_\phi$.
The "displacing" part of Eq. (\ref{outmz1}), together with the rotation 
on the mode $b$, can be easily rewritten as
\begin{eqnarray}
\exp\left\{i\frac{\pi}{2}b^\dag b\right\} 
\hat U_\phi \hat D_a (\alpha) \hat D_b (-i\alpha) \hat U_\phi^\dag 
=\hat D_a (\alpha e^{i\frac{\phi}{2}})\hat D_b (\alpha e^{\frac{i}{2}
(\pi-\phi)})\exp\left\{i\frac{\pi}{2}b^\dag b\right\} 
\label{displa}\;,
\end{eqnarray}
whereas the "squeezing" part needs a little more algebra: specializing 
a result from Ref. \cite{join} we can write 
\begin{eqnarray}
\hat U_\phi \hat S_a (r) \hat S_b (-r)\hat U_\phi^\dag = 
\exp\left\{\cos \phi \left[\frac{1}{2}r(a^{\dag 2} -a^2 - b^{\dag 2} + b^2)
\right] + \sin \phi \left[r(a^\dag b^\dag - ab)\right]\right\}
\label{sque}\;.
\end{eqnarray}
It is worth noting that squeezing at the input is essential to obtain
entanglement at the output. In fact, for the input state being a couple
of coherent states $|\psi_{\small \sc in}\rangle = \hat D_a (\alpha) 
\hat D_b (\beta) | {\bf 0} \rangle$ the output state is given by 
$|\psi_{\small \sc out}\rangle = \hat D_a (\alpha\cos\phi - i \beta\sin\phi) 
\hat D_b (-i\alpha\sin\phi+\alpha\cos\phi) | {\bf 0} \rangle$, which again 
is a couple of factorized (uncorrelated) coherent states for any value of the
internal phase-shift of the interferometer. Actually, the absence of output 
correlations is due the the Poissonian statistics of the coherent states, 
which implies the absence of intensity fluctuations \cite{man}.
\par
Let us first consider the situation $\phi=\frac{\pi}{2}$. In this case, the
transformation in Eq. (\ref{sque}) reduces to the two-mode squeezing 
operator $\hat S^{(2)} (r)= \exp\left\{r(a^\dag b^\dag - ab)\right\}$, so 
that the output state coincides with a displaced and  
rotated twin-beam state
\begin{eqnarray}
| \psi_{\small \sc out} \rangle = \hat D_a (\alpha e^{i\frac{\phi}{2}}) 
\hat D_b (\alpha e^{\frac{i}{2}( \pi - \phi)}) 
\exp\left\{i\frac{\pi}{2} b^\dag b\right\}|\psi_{\small \sc twb}\rangle 
\label{mzout2}\;,
\end{eqnarray}
where the explicit expression of the twin-beam state 
$|\psi_{\small \sc twb}\rangle$ is given by
\begin{eqnarray}
|\psi_{\small \sc twb}\rangle =\hat S^{(2)} (r) |{\bf 0}\rangle = 
\frac{1}{\cosh r} \sum_{k=0}^{\infty} \tanh^k r \: |k,k\rangle
\label{twb}\;.
\end{eqnarray}
In order to evaluate the degree of entanglement of $|\psi_{\small \sc out} 
\rangle$ we use the parameter $\epsilon$ introduced in Eq. (\ref{entdef}). 
The partial trace over a mode, say $b$, is given by 
\begin{eqnarray}
\hat\varrho_a=\hbox{Tr}_b \left\{
|\psi_{\small \sc out}\rangle\langle \psi_{\small \sc out}|  
\right\}= \frac{1}{\cosh^2 r} \sum_{k=0}^{\infty} \tanh^{2k} r \: 
\hat D(\alpha e^{i\frac{\pi}{4}}) |k \rangle\langle k | \hat 
D^\dag (\alpha e^{i\frac{\pi}{4}})
\label{rhoa}\;,
\end{eqnarray}
which is diagonal in the basis of displaced number states 
$|\psi_n\rangle=\hat D(\alpha e^{i\frac{\pi}{4}}) |n \rangle$. 
The set of $|\psi_n\rangle$'s constitutes an orthogonal basis for the 
Hilbert space of harmonic oscillator, and therefore the entropy 
$S[\hat\varrho_a]$ can be evaluated as 
\begin{eqnarray}
S[\hat\varrho_a] = - \sum_{n=0}^{\infty} p_n \log p_n \qquad
p_n=\frac{1}{\cosh^2 r} \tanh^{2n} r
\label{sa}\;.
\end{eqnarray}
After straightforward calculation we arrive at 
\begin{eqnarray}
S[\hat\varrho_a] = 
\log (1+ \nu^2) + \nu^2 \log (1+ \frac{1}{\nu^2})
\label{sa1}\;,
\end{eqnarray}
where $\nu^2=\sinh^2 r$ is the squeezing energy of each input beam. 
Notice that $S[\hat\varrho_a]$ in Eq. (\ref{sa1}) is equivalent to the entropy
of a thermal state with $\nu^2$ photons. The degree of entanglement is 
given by
\begin{eqnarray}
\epsilon = \frac{\log (1+ \gamma N) + \gamma N \log (1+ \frac{1}{\gamma N})}
{\log (1+ N) + N\log (1+ \frac{1}{N})}
\label{ent}\;,
\end{eqnarray}
where $N=\langle a^\dag a\rangle = |\alpha |^2 + \nu^2$ is the total 
energy of each input signal, and $\gamma$ is the squeezing fraction, 
namely the percentage of the total energy engaged in squeezing photons, 
$\nu^2=\gamma N$. 
From Eq. (\ref{ent}) it is apparent that the degree of entanglement is
an increasing function of the squeezing fraction, and that  
the maximum entangled state ($\epsilon=1$) at the output is reached for 
a couple of squeezed vacuum ($\gamma=1$) at the input. 
\par\noindent
For $\phi=0$, the transmissivity of the whole device is equal to unit,
and we have $|\psi_{\small \sc out}\rangle=|\psi_{\small \sc in}
\rangle$. Therefore, the entanglement is equal to zero, as 
the input state consists of a couple of uncorrelated signals.
\par\noindent
For $\phi\neq 0,\frac{\pi}{2}$ it is convenient to evaluate the output state 
and the entanglement by evolving the two-mode Wigner function, which is 
defined as follows
\begin{eqnarray}
W (x_a,y_a;x_b,y_b) &=& \int _{\mathbb R}\!\!d\mu_a \int_{\mathbb R}\!\!d\nu_a
\int _{\mathbb R}\!\!d\mu_b\int _{\mathbb R}\!\!d\nu_b \: \exp 
\left\{2i(\nu_a x_a -\mu_a y_a + \nu_b x_b - \mu_b y_b)\right\} \times
\nonumber \\ &\times& \hbox{Tr}\left\{\hat \varrho\:\hat D_a (\mu_a+ i\nu_a) 
\hat D_b (\mu_b + i \nu_b)\right\}
\label{wigdef}\;.
\end{eqnarray}
The $\pm \pi/2$ rotations of mode $b$ correspond to simple rotations in 
the sole $b$-variables
\begin{eqnarray}
\hat\varrho' &=& e^{i\frac{\pi}{2}} \hat\varrho \: e^{-i\frac{\pi}{2}}
\qquad \Longrightarrow  \qquad 
W' (x_a,y_a;x_b,y_b) = W (x_a,y_a;y_b,-x_b) \nonumber \\
\hat\varrho' &=& e^{-i\frac{\pi}{2}} \hat\varrho \: e^{i\frac{\pi}{2}}
\qquad \Longrightarrow  \qquad 
W' (x_a,y_a;x_b,y_b) = W (x_a,y_a;-y_b,x_b) 
\label{wigrot}\;,
\end{eqnarray}
whereas the action of the beam splitter $BS_\phi$, i.e. $\hat\varrho' = 
\hat U_\phi \hat\varrho \hat U_\phi^\dag$ corresponds to a mixing of 
variables of the two modes, in formula
\begin{eqnarray}
W' (x_a,y_a;x_b,y_b) = W 
(&& x_a \cos\delta -x_b\sin\delta,y_a  \cos\delta -y_b\sin\delta;\nonumber \\
 && x_a \sin\delta +x_b\cos\delta,y_a \sin\delta +y_b\cos\delta)
\label{wigbs}\;,
\end{eqnarray}
where we use the notation $\delta=\phi/2$.
Using Eqs. (\ref{wigrot}) and (\ref{wigbs}) the Wigner function
at the output results
\begin{eqnarray}
W_{\small \sc out}(x_a,y_a;x_b,y_b) = W_{\small \sc in} 
(&& x_a \cos\delta -y_b\sin\delta, y_a \cos\delta + x_b\sin\delta;\nonumber \\
&& x_b \cos\delta -y_b\sin\delta,x_a \sin\delta +y_b\cos\delta)
\label{wigout}\;,
\end{eqnarray}
where $W_{\small \sc IN} (x_a,y_a;x_b,y_b)$ is a product of two
identical single-mode Gaussian Wigner functions, corresponding to the
couple of input squeezed-coherent states: 
\begin{eqnarray}
W_{\small \sc in}(x_a,y_a;x_b,y_b) = \frac{4}{\pi^2} \exp\Bigg\{ 
&-&2 e^{-2r} (x_a-\hbox{Re} [\alpha ])^2 - 2 e^{2r}(y_a - \hbox{Im}[ 
\alpha ])^2 \nonumber \\ &-& 2 e^{-2r} (x_b-\hbox{Re}[ \alpha ])^2 
- 2 e^{2r}(y_b - \hbox{Im}[ \alpha ])^2 \Bigg\}
\label{wigin}\;.
\end{eqnarray}
By the integration over the $b$-variables
\begin{eqnarray}
W_{\small \sc out}(x_a,y_a) = \int _{\mathbb R}\!\!dx_b 
\int_{\mathbb R}\!\!dy_b\: W_{\small \sc out}(x_a,y_a;x_b,y_b)
\label{wigpar}\;,
\end{eqnarray}
and inserting Eqs.  (\ref{wigout}) and (\ref{wigin}) in Eq. (\ref{wigpar})
we obtain
\begin{eqnarray}
W_{\small \sc out}(x_a,y_a) = \frac{1}{\pi\Sigma_x\Sigma_y}
\exp\left\{-\frac{(x_a-\hbox{Re}[\alpha_{\phi}])^2}{\Sigma_x^2}-
\frac{(y_a-\hbox{Im}[\alpha_{\phi}])^2}{\Sigma_y^2}\right\} 
\label{wigone}\;,
\end{eqnarray}
which represents the Wigner function of the sole mode $a$ after partial trace
over the mode $b$. The quantities $\Sigma_x$ and $\Sigma_y$ in Eq.
(\ref{wigone}) are given by 
\begin{eqnarray}
\Sigma_x^2 &=& e^{2r}\cos^2\delta+e^{-2r}\sin^2\delta\nonumber \\
\Sigma_y^2 &=& e^{-2r}\cos^2\delta+e^{2r}\sin^2\delta
\label{sigmas}\;,
\end{eqnarray}
whereas $\alpha_{\phi}$ is given by
\begin{eqnarray}
\alpha_{\phi}=\alpha\sqrt{1+\frac12\sin^2\phi}
\label{alp}\;.
\end{eqnarray}
In order to evaluate entanglement, we note that any unitary transformation
$\hat T$ acting on the single mode $a$ does not change the value of the 
entropy \cite{bern}, i.e.  
$S[\hat\varrho_a] = S[\hat T \hat\varrho_a \hat T^\dag]$. 
Using this property, we displace with amplitude $ \alpha_\phi$, and 
then squeeze with parameter $r^\ast=\log \sqrt{\Sigma_y/\Sigma_x}$ the 
Wigner function in Eq. (\ref{wigone}), thus arriving at the following 
entropy-equivalent state
\begin{eqnarray}
W'_{\small \sc out}(x_a,y_a) = \frac{1}{\pi\Sigma_x\Sigma_y}
\exp\left\{-\frac{x_a^2+y_a^2}{\Sigma_y\Sigma_x}\right\} 
\label{wigth}\;.
\end{eqnarray}
Remarkably, the Wigner function in Eq. (\ref{wigth}) coincides with the 
Wigner function of a thermal states with thermal photons given by 
\begin{eqnarray}
N_\phi = \frac{1}{2}\left[\Sigma_y\Sigma_x - 1\right]
= \frac{1}{2}\left[\sqrt{1+ \sin^2\phi\sinh^2 2r} 
-1\right]\label{nth}\;.
\end{eqnarray}
The corresponding entropy can be easily computed, and thus the entanglement 
at the output is given by 
\begin{eqnarray}
\epsilon = \frac{\log (1+ N_\phi) + N_\phi \log (1+ \frac{1}{N_\phi})}
{\log (1+ N) + N\log (1+ \frac{1}{N})}\label{entphi}\;.
\end{eqnarray}
As it is expected, one has $N_\phi=0$ for $\phi=0$, and 
$N_\phi= \gamma N$ for $\phi=\pi/2$.  
\par
\begin{figure}
\begin{tabular}{cc}
\psfig{file=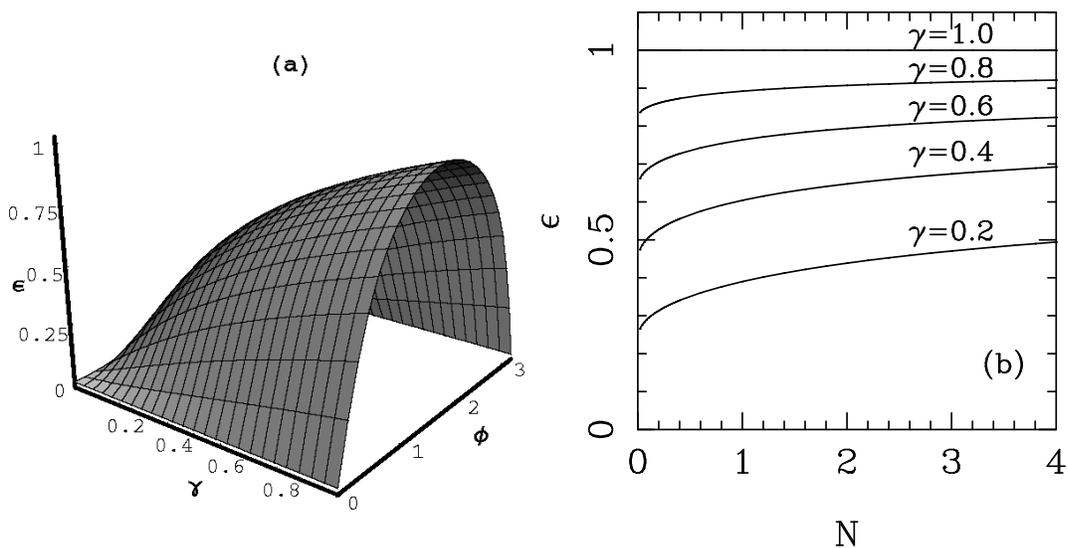,width=7cm}
 & \psfig{file=fig2b.ps,width=7cm}
\end{tabular}
\caption{(a):the degree of entanglement as a function of the squeezing 
fraction $\gamma$ and the internal phase-shift $\phi$, in the case of 
input beams with $N=3$ each: for fixed $\gamma$ the output state ranges 
from a totally disentangled state for $\phi=0$, to a state with a degree 
of entanglement given by Eq. (\ref{ent}) for $\phi=\frac{\pi}{2}$. 
The degree of entanglement is an increasing function of $\gamma$,
with $\phi=\frac{\pi}{2}$ corresponding to maximum value.  
Different values of $N$ do not substantially modifies the picture.
(b): the degree of entanglement as a function of the total input
energy $N$ for different values of the squeezing fraction $\gamma$, and 
for fixed value $\phi=\frac{\pi}{2}$ of the internal phase-shift. 
\label{f:ent}}
\end{figure}

\begin{figure}
\psfig{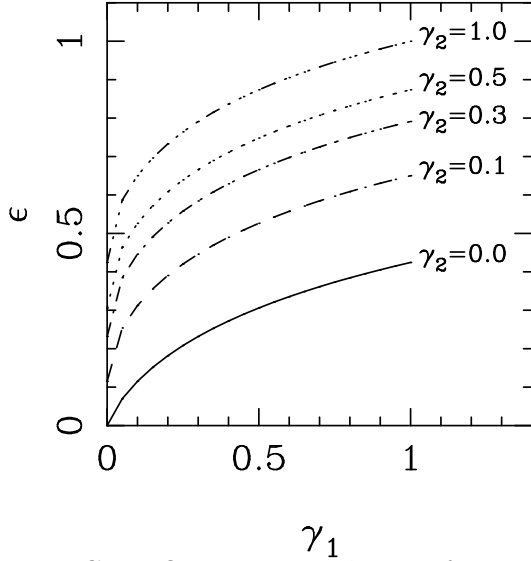}
\caption{Output entanglement for input signals with different degree of
squeezing. The maximum entanglement at the output (for $\phi=\pi/2$) is 
reported as a function of the squeezing fractions $\gamma_1$ of one of 
the beams for different values of the squeezing fraction $\gamma_2$ 
of the other beam. Both input beams have an average number of photons 
equal to $N=3$. The output entanglement is an increasing function of both 
the two squeezing fractions. The extreme case in which one 
of the input signals is no squeezed at all corresponds to a value of 
$\epsilon$ always lower than $50\%$. \label{f:tre}}
\end{figure}

 
In Fig. \ref{f:ent}a we show the degree of entanglement
as a function of the squeezing fraction $\gamma$ and the internal 
phase-shift $\phi$, in the case of input beams with average photons 
$N=3$ each: at fixed $\gamma$ the output state ranges from a totally 
disentangled state for $\phi=0$, to a state whose degree of entanglement 
is given by Eq. (\ref{ent}) for $\phi=\frac{\pi}{2}$. The degree of 
entanglement 
is an increasing function of the squeezing fraction $\gamma$,
with the condition $\phi=\frac{\pi}{2}$ corresponding to maximum value.  
Different values of the intensity $N$ does not substantially modify 
the behavior of $\epsilon$ versus $\gamma$ and $\phi$.
In Fig. \ref{f:ent}b we report $\epsilon$ as a function of the intensity 
$N$ for different values of the squeezing fraction $\gamma$, 
and for fixed value $\phi=\frac{\pi}{2}$ of the internal phase-shift:
For $\gamma=1$ one has $\epsilon=1$ independently on $N$, whereas for $\gamma
<1$ the degree of entanglement becomes a slightly increasing function of $N$.
For highly excited states the entanglement is given by the asymptotic 
formula 
\begin{eqnarray}
\epsilon \stackrel{N \gg 1}{\simeq} 1+ \frac{\log\gamma}{\log N}
\label{entasy}\;.
\end{eqnarray}
So far we have considered the two input states having the same degree of 
squeezing. However, a pair of input states with different squeezing fractions 
does not substantially modify the picture. In this case, in fact, the 
entanglement still oscillates from $\epsilon=0$ to a maximum
value as a function of the internal phase-shift of the interferometer. 
On the other hand, this maximum value is now a function of both the squeezing
fractions, and maximally entangled states at the output cannot be achieved 
if one of the input signals is only partially squeezed. In Fig. \ref{f:tre} 
we report the maximum entanglement at the output (obtained for $\phi=\pi/2$) 
as a function of the squeezing fractions $\gamma_1$ of one of the beams for 
different values of the squeezing fraction $\gamma_2$ of the other beam.
The plots refers to a situation in which both input beams have an average 
number of photons equal to $N=3$. As it is apparent from the plots, the 
output entanglement is an increasing function of both the two squeezing 
fractions. The extreme case in which one 
of the input signals is no squeezed at all corresponds to a value of 
$\epsilon$ always lower than $50\%$. 
\section{Entanglement and fringes visibility}\label{s:vis}
In this Section, we study the visibility of the interference fringes that 
are observed, by varying the internal phase-shift $\phi$, in intensity 
measurements at the output of the interferometer.  
In analogy with experiments involving correlated photon pairs, we consider 
the detection of the coincidence counting rate at the output, namely of the
fourth-order correlation function $\langle\psi_{\small \sc out}| 
a^\dag a \: b^\dag b|\psi_{\small \sc out}\rangle$. However, as we will show in 
the following, this corresponds to low fringes visibility, and thus we sought
for a more sensitive kind of measurement. 
The homodyne-like detection of the output difference photocurrent 
$\langle\psi_{\small \sc out}| a^\dag a - b^\dag b|\psi_{\small \sc 
out}\rangle$ is widely used in interferometry \cite{cav,bon,par}, 
and generally results 
in a very sensitive measurement scheme. Starting from this 
consideration, we suggest the squared difference photocurrent 
$\langle\psi_{\small \sc out}| (a^\dag a - b^\dag b)^2 |\psi_{\small \sc out}
\rangle$ as a suitable fourth-order quantity to be measured at the output of
the interferometer. 
\par
Besides being originated by interference effects, the variations in the
quantities measured at the output also reflect the variations in the quantum
correlations between the two output signals. Therefore, the visibility
of the interference fringes provides a measure of entanglement, and 
comparing the visibility of different measurement schemes 
provides a way to compare their ability in monitoring the variations of 
quantum correlations between the output signals.
\par
\begin{figure}
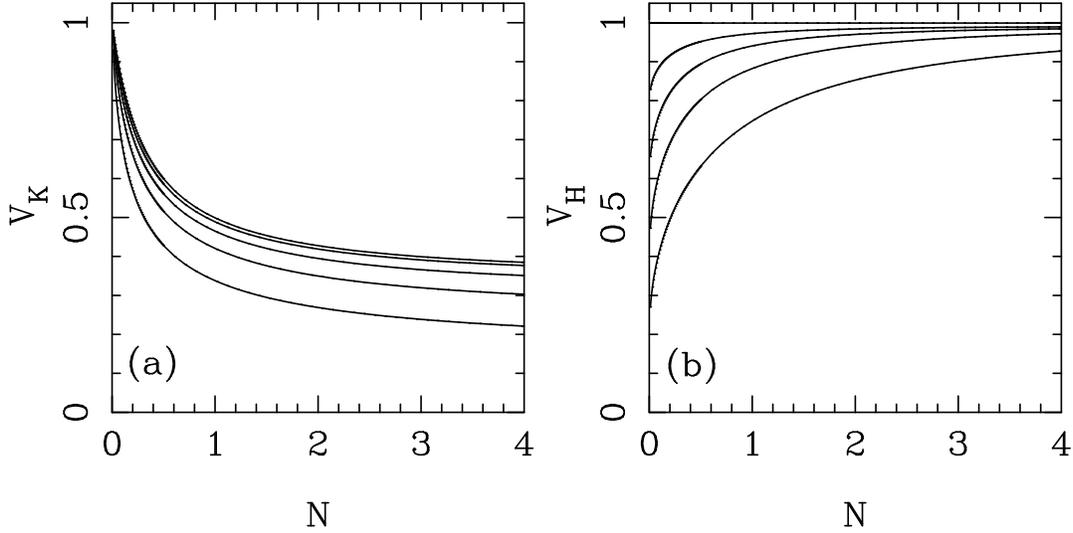

\begin{tabular}{cc}
\psfig{file=fig4a.ps,width=7cm}
 & \psfig{file=fig4b.ps,width=7cm}
\end{tabular}
\caption{Fringes visibility as a function 
of the intensity $N$ for different values of the input squeezing fraction
$\gamma$. In (a) the visibility of K-measurement $V_K$, and in (b) the 
visibility of H-measurement $V_H$ . 
In both plots we report the visibility versus $N$ for five values of the
input squeezing fraction. From bottom to top we have the curves for
$\gamma=0.2,0.4,0.6,0.8, \hbox{and } 1.0$. 
As it is apparent, $V_H$ is larger than 
$V_K$ in almost all situations, with the exception of the very low 
signals regime. 
\label{f:vis}}
\end{figure}

As already mentioned, here we consider the measurement of the coincidence
counting rate 
\begin{eqnarray}
K(\phi) =\langle\psi_{\small \sc out}| a^\dag a \: b^\dag b|\psi_{\small 
\sc out}\rangle = \langle\psi_{\small \sc in}|\hat V_{\small MZ}^\dag (\phi) 
a^\dag a \: b^\dag b\:\hat V_{\small MZ} (\phi)|\psi_{\small \sc in}\rangle
\label{kdef}\;,
\end{eqnarray}
and of the squared difference photocurrent
\begin{eqnarray}
H (\phi)=\langle\psi_{\small \sc out}| (a^\dag a - b^\dag b)^2|\psi_{\small 
\sc out}\rangle = \langle\psi_{\small \sc in}|\hat V_{\small MZ}^\dag (\phi) 
(a^\dag a -b^\dag b)^2\:\hat V_{\small MZ} (\phi)|\psi_{\small 
\sc in}\rangle\label{hdef}\;.
\end{eqnarray}
After some algebra, we arrive at the explicit expressions
in terms of the input fields
\begin{eqnarray}
\hat K (\phi) &=& \hat V_{\small MZ}^\dag (\phi) 
a^\dag a \: b^\dag b\:\hat V_{\small MZ} (\phi) =\nonumber \\ 
&=&\sin^2\delta \cos^2\delta 
\left[a^{\dag 2}a^2 + b^{\dag 2}b^2+a^{\dag 2}b^2+b^{\dag 2}a^2\right] 
+ (\sin^2\delta - \cos^2\delta)^2 a^\dag a \: b^\dag b  + \nonumber \\
&+& i \sin\delta \cos^3\delta \left[
ab^{\dag 2} b+a^{\dag 2} a b-a^\dag b^\dag b^2-a^\dag a^2 b^\dag
\right] + \nonumber \\ 
&+& i \sin^3\delta \cos\delta \left[a^\dag a^2 b^\dag+ 
a^\dag b^\dag b^2 -a^{\dag 2} a b -a b^{\dag 2} b \right]
\label{kexp}\;,
\end{eqnarray}
\begin{eqnarray}
\hat H (\phi) &=& \hat V_{\small MZ}^\dag (\phi) (a^\dag a - b^\dag b)^2 
\:\hat V_{\small MZ} (\phi) = - 2 \hat K (\phi) + \nonumber \\
&+& \left[(a^\dag a)^2+(b^\dag b)^2\right] (\sin^4 \delta + \cos^4 \delta) -
2 \sin^2\delta \cos^2\delta\left[a^{\dag 2}b^2 + b^{\dag 2}a^2 -a^\dag a-
b^\dag b\right] + \nonumber \\&+& 2i \sin\delta \cos^3\delta 
\left[a^\dag a^2 b^\dag - a^{\dag 2} a b +a^\dag b^\dag b^2 -
a b^{\dag 2} b \right] + \nonumber \\ &+& 2i \sin^3\delta \cos\delta \left[
a b^{\dag 2} b - a^\dag b^\dag b^2 + a^{\dag 2} a b- a^\dag a^2 b^\dag \right]
\label{hexp}\;,
\end{eqnarray}
where again we used the notation $\delta=\frac{\phi}{2}$.
Using Eqs. (\ref{kexp}) and (\ref{hexp}) we are able to evaluate the
fringes visibility of both detection schemes
\begin{eqnarray}
V_K =\frac{K_{max}-K_{min}}{K_{max}+K_{min}} \qquad
V_H =\frac{H_{max}-H_{min}}{H_{max}+H_{min}}
\label{hkvis}\;.
\end{eqnarray}
In Fig. \ref{f:vis} we report $V_K$ and $V_H$ as a function 
of the intensity $N$ for different values of the input squeezing fraction
$\gamma$. The H-measurement visibility $V_H$ is larger than 
$V_K$ in almost all situations, with the exception of the very low 
signals regime, where very few photons are present. 
The behavior of fringes visibility versus intensity $N$ also confirms that 
$V_H$ represents a good measure of the entanglement at the output. 
As it happens for the degree of entanglement, in fact, a couple of 
squeezed vacuum at the input corresponds to maximum visibility $V_H=1$ 
independently on the intensity,. On the other hand, the coincidence 
counting rate shows a visibility $V_K$ that rapidly decreases versus 
$N$, and saturates to a value well below $1/2$. 
For non unit squeezing fraction, and moderate input intensities $N <10$, 
the behavior of $V_H$ looks qualitatively similar to that of the degree of 
entanglement (compare Fig. \ref{f:vis}b and Fig. \ref{f:ent}b), whereas
again $V_K$ rapidly decreases. Remarkably, for highly excited
states $N >10$, the visibility $V_H$ has the same asymptotic dependence 
of the degree of entanglement $\epsilon$, in formula
\begin{eqnarray}
\epsilon \stackrel{N \gg 1}{\simeq} 1+ \frac{A(\gamma)}{\log N}
\label{visasy}\;,
\end{eqnarray}
where the proportionality constant $A(\gamma)\simeq 1/5 \log\gamma$ is 
roughly proportional to that appearing in Eq. (\ref{entasy}). 
\section{Conclusions}\label{s:outro}
The generation and the detection of optical entangled states are important 
issues, both required for fundamentals tests of quantum mechanics, as well as 
for possible applications.
In this paper we have studied the entanglement between the two beams
exiting a Mach-Zehnder interferometer fed by a couple of squeezed-coherent 
states with arbitrary squeezing parameter. 
The degree of entanglement at the output has been analytically evaluated, as 
a function of the input intensity and squeezing fraction, and of the internal
phase-shift of the interferometer. Our results indicate that 
entangled states of arbitrary large intensity can be produced by varying 
the input energy, whereas the degree of entanglement can be tuned by 
varying the input squeezing fraction, and the internal phase-shift. \par
An experimental characterization of the output entanglement can be obtained 
through the measurement of the squared difference photocurrent between the
output modes. The interference fringes that are observed by varying the 
internal phase-shift $\phi$ show, in fact, high visibility for the whole 
range of input squeezing parameter. 
\section*{Acknowledgments}
The author would thank Valentina De Renzi for valuable discussions 
and the ``{\em Accademia Nazionale dei Lincei} '' for 
financial support through the ``{\em Giuseppe Borgia} '' award. 


\begin{thebibliography}{99}
\bibitem{note} Our definition slightly differs from that of Refs. 
\cite{bp91,lind,bp89}, where no normalization to a thermal state was used.
\end{thebibliography}
\end{document}